\documentclass[sigconf]{acmart}
\AtBeginDocument{%
  }

\copyrightyear{2025}
\acmYear{2025}
\setcopyright{acmlicensed}\acmConference[Websci '25]{Proceedings of the 17th ACM Web Science Conference 2025}{May 20--24, 2025}{New Brunswick, NJ, USA}
\acmBooktitle{Proceedings of the 17th ACM Web Science Conference 2025 (Websci '25), May 20--24, 2025, New Brunswick, NJ, USA}
\acmDOI{10.1145/3717867.3717899}
\acmISBN{979-8-4007-1483-2/2025/05}

\begin{document}

\title{Personal Narratives Empower Politically Disinclined Individuals to Engage in Political Discussions}

\author{Tejasvi Chebrolu}
\authornote{Work done while at UT Austin}
\email{tejasvi.chebrolu@research.iiit.ac.in}
\orcid{0009-0007-1441-0325}
\affiliation{%
  \institution{International Institute of Information Technology, Hyderabad}
  \city{Hyderabad}
  \state{Telangana}
  \country{India}
}

\author{Ponnurangam Kumaraguru}
\email{pk.guru@iiit.ac.in}
\orcid{0000-0001-5082-2078}
\affiliation{%
  \institution{International Institute of Information Technology, Hyderabad}
  \city{Hyderabad}
  \state{Telangana}
  \country{India}
}

\author{Ashwin Rajadesingan}
\authornote{Corresponding author}
\email{arajades@austin.utexas.edu}
\orcid{0000-0001-5387-1350}
\affiliation{%
  \institution{University of Texas at Austin}
  \city{Austin}
  \state{Texas}
  \country{USA}
}

\begin{abstract}
  Engaging in political discussions is crucial in democratic societies, yet many individuals remain politically disinclined due to various factors such as perceived knowledge gaps, conflict avoidance, or a sense of disconnection from the political system. In this paper, we explore the potential of personal narratives—short, first-person accounts emphasizing personal experiences—as a means to empower these individuals to participate in online political discussions. Using a text classifier that identifies personal narratives, we conducted a large-scale computational analysis to evaluate the relationship between the use of personal narratives and participation in political discussions on Reddit. We find that politically disinclined individuals (PDIs) are more likely to use personal narratives than more politically active users. Personal narratives are more likely to attract and retain politically disinclined individuals in political discussions than other comments. Importantly, personal narratives posted by politically disinclined individuals are received more positively than their other comments in political communities. These results emphasize the value of personal narratives in promoting inclusive political discourse.
\end{abstract}



\maketitle

\section{Introduction}
Political discussions between ordinary citizens are central to a democratic society. Such discussions result in a wide range of normatively desired outcomes such as increased political knowledge \cite{eveland2006talking}, informed opinion formation \cite{cappella2002argument} and political tolerance \cite{mutz2002cross}. Yet, most individuals do not engage in political discussions, especially online. For example, in surveys with American participants, conducted by the Pew Research Center, one in three respondents never discussed politics with their friends \cite{jurkowitz2020sore}, and 70\% of Americans rarely or never post about political content on social media \cite{pewresearchUSSocial}. 

Curiously, not everyone refrains from engaging in political discussions. Krupnikov and Ryan \cite{krupnikov2022other} find that some individuals are ``deeply involved'' in politics and are often very vocal in political discussions. These individuals are a hyperpartisan, affectively polarized minority group for whom politics is a central part of their lives. These individuals typically engage in hostile interactions with both outparty supporters and those within the party who they perceive as not being sufficiently extreme \cite{bail2022breaking}. Although a small minority, these deeply involved individuals dominate online political discussions, making them the most visible to others \cite{krupnikov2022other}. Journalists, who often disproportionately focus their reporting on extreme partisans, further amplify this dynamic, creating an exaggerated perception of political polarization \cite{levendusky2016mis,krupnikov2022other}.

How best to address this issue of the deeply involved dominating political discourse online? This insularity can create echo chambers that deepen political polarization and diminish openness to opposing viewpoints \cite{justwan2018social}. It also marginalizes less politically active individuals, further narrowing the diversity of voices in discourse and ultimately weakening democratic engagement. The most straightforward solution is to encourage others to engage with politics online \cite{bail2022breaking}. The problem with this solution is that, while the majority of other users are less polarized and more moderate, they are also much less vocal about politics \cite{krupnikov2022other}. How then can we facilitate participation from this politically reclusive group? We argue that encouraging the use of personal narratives in political discussions may facilitate greater participation among those who are disinclined towards politics. Recent research suggests that apart from concerns about maintaining social ties, a lack of political knowledge or understanding was cited as a major reason for avoiding political discussions \cite{carlson2022goes}. Personal narratives may address this concern by allowing individuals to rely on their own personal experiences when engaging with politics. Further, personal narratives and storytelling can foster more inclusive political communication by presenting situated knowledge and enabling disadvantaged groups who may lack access to formal argumentative skills to effectively share their perspectives in the public sphere \cite{young2002inclusion}. Similarly, Polletta points to how storytelling can be equalizing in deliberative settings where participants have different knowledge and argumentation skills, ``since everyone has his or her own story'' \cite{polletta2009like}.

To evaluate how personal narratives can help individuals engage in political discussions, we conducted a large-scale computational analysis of public comments in political subreddits on Reddit from 2020 to 2021. We identified two groups of users that we are most interested in: users who have rarely engaged and users who have not at all engaged in discussions in political subreddits in the previous 12 months. We formally defined these two groups as politically disinclined individuals (PDIs). We compared these users with users who are most active in these political communities. Then, we fine-tuned a transformer-based classifier to identify personal narratives in comments posted in political subreddits. Across multiple analyses, we found strong evidence linking the use of personal narratives and participation in political discussions.

First, we found that PDIs were more likely to use personal narratives in their comments in political subreddits than more politically active users. Further, comments containing personal narratives were slightly more likely to receive responses from PDIs than other comments in political subreddits. Our analysis also revealed that PDIs who employed personal narratives in a given month were more likely to continue participating in political discussions in the subsequent month compared to those who did not use a personal narrative. These results suggest that personal narratives not only attract the politically disinclined but also encourage their sustained involvement in political discourse over time.

Importantly, personal narratives appear to be positively viewed by the political community members themselves. Reddit's karma scoring system allows members to upvote or downvote comments. Thus, a higher score implies that the comment is positively viewed by the community. We find that personal narratives, on average, garner higher scores than other comments by PDIs. Although there appears to be a perception gap, that is, comments by more politically active users, on average, score higher than PDIs, personal narratives appear to significantly reduce this difference. Overall, these results present an encouraging picture of how personal narratives may facilitate participation among the politically disinclined.

\begin{table*}[ht]
\centering
\caption{Examples of personal narratives comments in the dataset.}
\label{tab:personal_narratives}
\resizebox{\textwidth}{!}{%
\begin{tabular}{|p{\textwidth}|}
\hline
\textbf{Example comments} \\ 

\hline
I'm a cdl-a truck driver in the USA. I work 60 hours over 5 days. Teamster contract ensures that I\'m well paid but we are paid hourly. A reduction in hours would cause hardship and with the shortage of drivers, a shortage of goods \\ 
\hline
I'm in Alabama and oh my god it was so humid yesterday. I was so unproductive from how bad it was \\ 
\hline
Buf you personally... How much is that cost? Cuz I got \$500 a month coming out of my pay, then deductibles, then coverage issues, not to mention dental; vision... Oh; ya know what's not afforded in that any kind of mental health coverage... \\ 
\hline
I faked being a liberal for my years at university. They never suspected a thing. \\ 
\hline
Due to circumstances not within my control, my (gender nonconforming) partner and i have been at my fathers these past few months. My partner has been forced to present as male because he knows we're in a house full of ultra conservatives ... \\ 
\hline
\end{tabular}%
}
\end{table*}

\section{Related work}

\subsection{Avoiding political discussions}

Over decades, scholars have suggested different explanations for why individuals may avoid engaging in political discussions. Eliasoph in \textit{Avoiding politics} argues that rather than an aversion to politics, lack of political talk is more because of a fragile public sphere where ``intelligence, curiosity, and generosity have evaporated'' \cite{eliasoph1998avoiding}. Noelle-Neumann’s spiral of silence theory suggests that individuals avoid expressing their opinions when they are perceived to be unpopular \cite{noelle1974spiral}. Others attribute avoidance to individual predispositions such as conflict avoidance \cite{ulbig1999conflict} and the Big Five personality traits \cite{hibbing2011personality}. Hostility \cite{matthes2013hostile} and disagreement \cite{gerber2012disagreement} have also been identified as factors in political discussion avoidance. More recently, through a series of studies, Carlson and Settle highlight the importance of political knowledge in the decision to not engage in political discussions \cite{carlson2022goes}. Specifically, in one study, they find that one of the most commonly cited reasons for political discussion avoidance was the respondent's lack of accurate political information or knowledge \cite{carlson2022goes}.

While a large majority avoid or only rarely engage in political discussions, a small but vocal minority of individuals who are deeply involved in politics dominate political discussions \cite{krupnikov2022other}. These individuals are highly politically engaged and are often extreme partisans who are hostile towards those they disagree with \cite{krupnikov2022other}. Thus, most political discussions that we observe online are invariably between these deeply involved individuals. A public sphere with largely the deeply involved engaging in discussions is especially problematic. Interactions predominantly between deeply involved individuals are likely to result in more extreme views and more polarization \cite{sunstein2001republic}. Further, the visibility of deeply involved individuals engaging in hostile discussions, coupled with an increased media focus on them, often distorts perceptions of polarization and who is the median partisan \cite{levendusky2016mis,bail2022breaking,krupnikov2022other}. Finally, as Berelson et al. \cite{berelson1954voting} in their foundational work on opinion formation note, a mass democracy cannot function if all individuals are deeply involved. A wider distribution of individuals based on political involvement may facilitate compromise, avert extreme partisanship and provide room for consensus and stability in democratic decision-making \cite{berelson1954voting}. Thus, it is crucial that political discussions include individuals who do not usually engage with politics. We argue that personal narratives can facilitate a wider engagement among the politically disinclined.

\subsection{Personal narratives in political discussions}

Personal narratives are usually first-person accounts that recount individual experiences. Engaging with personal narratives requires minimal prior knowledge or training, allowing individuals from diverse backgrounds to participate meaningfully in political discourse \cite{young2002inclusion}. Personal narratives also make political discussions more accessible and create a safe space for dialogue, encouraging participation even among those hesitant to express their views publicly \cite{patterson1998narrative}. Moreover, personal narratives help situate individual stories within the context of larger political systems, enabling individuals to see how their experiences relate to broader societal issues \cite{riessman2003analysis}. Political apathy, another key reason for political inactivity, can also be countered through personal narratives as they create connections across ideological differences, encourage engagement via relatable storytelling, and highlight the impact of individual experiences within broader political and social systems \cite{jones2012contesting}. Finally, personal narratives help build inclusive spaces by emphasizing individual experiences as valid forms of knowledge, which is crucial for engaging PDIs who may feel excluded from traditional political discourses \cite{stivers1993reflections}. 

Personal narratives may positively impact the substance of the political discussions as well. These narratives often humanize the narrator, either by showcasing their positive qualities or revealing their vulnerabilities, creating a sense of connection with readers \cite{robinson1981personal}. This connection is further strengthened by the empathy personal narratives can evoke, making readers more open to considering opposing viewpoints \cite{shuman2006entitlement}. Personal narratives can also inspire political action by turning personal stories into powerful tools for raising awareness and building solidarity \cite{robinson2016sharing}. Similarly, employing these narratives in political discourse also grants users tools for persuasion in political contexts, as these narratives have the ability to influence attitudes and beliefs, unlike data-heavy arguments, as stories often appeal directly to emotions \cite{green2000role}. Somewhat counterintuitively, personal narratives also enhance perceptions of rationality in political discussions. When people use personal narratives to express their political views, they are often seen as more rational, garnering greater respect even from opposing partisan groups \cite{kubin2021personal}.  Finally, personal narratives can amplify the voices of underrepresented groups, challenge mainstream narratives, and inspire hope while advocating for change \cite{10.1093/oso/9780190851712.003.0004}. These studies suggest that personal narratives may not only encourage PDIs to participate in political discussions but may also substantively improve the quality of the discussions.

\subsection{Identifying personal narratives}

Research on identifying and extracting personal narratives from textual data has evolved over the years. Gordon and Swanson laid the groundwork by creating a standard corpus of personal narratives from blog posts and using statistical models to classify them \cite{gordon2009identifying}. Subsequent research explored various statistical methods for narrative identification, focusing on different linguistic characteristics. Yao and Huang examined temporal characteristics \cite{yao2018temporal}, while Ceran et al. investigated the density of part-of-speech tags and named entities \cite{ceran2012semantic}.
Researchers then expanded their approaches to incorporate semantic information. Eisenberg and Finlayson utilized verb and character features \cite{eisenberg2017simpler}, and Dirkson et al. leveraged psycholinguistic features to improve narrative identification \cite{dirkson2019narrative}. 
Recent studies have demonstrated the potential of transformer-based models in personal narrative identification. Ganti et al. conducted a study where they documented the performance of various transformer-based models \cite{ganti2023narrative}, while Anotoniak et al. successfully fine-tuned these models to identify narratives at both document and span levels across different domains \cite{antoniak2023people}. Falk and Lapesa further validated the robustness of transformer-based models in identifying narratives in argumentation settings \cite{falk-lapesa-2022-reports}. Generative models, including large language models,  have also been used to identify personal narratives \cite{ganti2023narrative}, but they performed worse than fine-tuning a transformer-based model. 

\section{Hypotheses}
We evaluate how personal narratives may affect political engagement. As discussed earlier, individuals often feel disenfranchised in political discourse, often due to a perceived lack of sufficient political knowledge to contribute meaningfully \cite{carlson2022goes}. However, personal narratives can help overcome this barrier by reframing topics into relatable personal stories, offering unique perspectives that make the subject matter less intimidating and more accessible \cite{lea2005understanding}. Similarly, the approachability of personal narratives likely engages less politically active users more than other kinds of content. Therefore, we hypothesize the following:
\begin{quote}
    \textbf{H1:} PDIs are more likely to use personal narratives than the most politically active users.
    
    \textbf{H2:} Personal narratives are more likely to receive responses than other comments from PDIs.
\end{quote}

Personal narratives may also foster a sense of community and belonging, creating bonds that encourage sustained participation \cite{rappaport1993narrative, lea2005understanding}. Based on this, we hypothesize that:
\begin{quote}
    \textbf{H3:} PDIs who use personal narratives in political communities are more likely to engage in political communities the following month than those who do not use personal narratives. 
\end{quote}

\begin{table*}[h!]
\renewcommand{\arraystretch}{1}
\centering
\caption{Mean performance metric per fold for the personal narrative classifier. The classifier achieves a macro average f-1 score of 0.82.}
\label{Table 1}
\begin{tabular}{|c|c|c|c|c|}
\hline
Metric & Not Personal Narrative & Personal Narrative & Macro Average & Weighted Average \\
\hline
Precision & 0.8425 & 0.7986 & 0.8288 & 0.8302 \\
Recall    & 0.8716 & 0.7728 & 0.8147 & 0.8325 \\
F1 Score  & 0.8446 & 0.8032 & 0.8211 & 0.8219 \\
Support   & 231.6  & 168.0  & 400.0  & 400.0  \\
Accuracy  & -      & -      & -      & 0.8320 \\
\hline
\end{tabular}
\end{table*}

Next, we evaluate community perceptions of personal narratives in political discussions. On Reddit, comments can be upvoted or downvoted, with each upvote increasing a comment's score by one point. A higher score indicates a positive reception by the community compared to comments with lower scores. Given the PDIs limited experience and engagement with political communities compared with highly politically active users, we expect that their contributions are likely less valued in political communities. Therefore, we expect that:

\begin{quote}
    \textbf{H4:} Comments by PDIs are likely to score lower than comments by the most politically active users.
\end{quote}

Personal narratives often provide new information or viewpoints that are unique to the discussion. This novelty in personal narratives is often valued in user-contributed comment sections. For example, the New York Times explicitly chooses to highlight personal stories in their comments section. \footnote{\url{https://archive.nytimes.com/www.nytimes.com/times-insider/2014/04/17/a-comments-path-to-publication/}} Therefore, we hypothesize that:
\begin{quote}
    \textbf{H5:} Personal narrative comments by politically disinclined individuals are more likely to score higher than other comments by PDIs.
\end{quote}

Finally, if H4 is true, we evaluate if personal narratives can bridge the perception gap, that is, the difference in positive perceptions of content posted by PDIs and most politically active users. We pose the following research question:
\begin{quote}
    \textbf{RQ1:} Can personal narratives reduce the perception gap between content posted by PDIs and most politically active users?
\end{quote}

\section{Data}
We conducted our analysis on public Reddit comments from 2020 to 2021, utilizing the Pushshift data archive \cite{baumgartner2020pushshift}. We used Rajadesingan et al.'s \cite{rajadesingan2021political} 2020 classification of political subreddits derived from a large-scale analysis of the prevalence of political talk in subreddits. In total, we identified 483 political subreddits, manually removing communities that were around mock/model elections, and adding  covid-related communities which were created in 2021 as the pandemic was highly politicized \cite{druckman2021affective}. In total, our dataset comprised 26,285,619 users who participated in these political communities. To ensure data quality, we excluded moderator and bot accounts. The dataset and code is available on GitHub. \footnote{\url{https://github.com/Ashwin-R/WebScience25-Personal-Narratives}}

\section{Methodology}

\subsection{Identifying personal narratives}

We define personal narratives as simply accounts of personal experiences, diverging from the traditional Labovian paradigm which emphasizes structured narratives with a clear beginning, middle, and end, centered on specific events, places, and times \cite{labov1972language}. We do this as, narrative analysis research, particularly in the context of social media, shows that stories told in these platforms are ``small stories'' that capture ordinary and mundane events from everyday life \cite{georgakopoulou2017small}. These stories often do not adhere to a traditional narrative structure, omitting the beginning or leaving the end to be filled in by others \cite{bamberg2008small}. Given the size of our dataset, we employed an automated approach to identify personal narratives.

\subsubsection{Personal Narrative Classifier}

Our approach builds on previous computational work that identifies personal narratives in online discussions. Specifically, both Antoniak et al. \cite{antoniak2023people} and Falk and Lapesa \cite{falk-lapesa-2022-reports} fine-tuned BERT-based classifiers on discussion data to identify personal narratives. In a pilot experiment, we found that the Falk and Lapesa's classifier \cite{falk-lapesa-2022-reports} performed better on our dataset likely because their broad conception of personal reports better aligns with our operationalization of personal narratives. We further fine-tuned this model to use in our dataset.

To generate additional training data, we ran their classifier on a sample of 20,000 comments which contained first-person pronouns (e.g., ``I'', ``we'') as potential markers of personal stories. We also filtered for comments with at least 50 characters as the pilot study revealed that shorter comments rarely contained personal narratives. From this sample, we randomly selected 1,000 comments classified by the model as personal narratives and 1,000 classified as non-narratives. Two trained annotators manually labeled these according to our operational definition of personal narratives. Initial annotations on a subset of 1,000 samples achieved an inter-rater reliability score of 0.79 (Cohen’s $\kappa$), after which one annotator completed the remaining labeling, yielding a final dataset of 2,000 samples (842 personal narratives, 1158 other comments) for fine-tuning.

This approach allowed us to create a balanced corpus without manually labeling large volumes of data—a challenging task given the relative rarity of personal narratives \cite{gordon2009identifying}. Our method also aligns with pseudo-labeling strategies that address the need for large labeled datasets \cite{rizve2021defense}. We then fine-tuned  \citep{falk-lapesa-2022-reports}'s classifier model on the new training set, training the model with a batch size of 32 over 10 epochs. The AdamW optimizer \cite{loshchilov2017decoupled} was used, with a learning rate of $5e-5$, and model evaluation was conducted via 5-fold cross-validation. Table \ref{Table 1} shows average evaluation metrics from these folds.

To evaluate the classifier's generalizability, we tested it on a random sample of 10,000 unseen comments. From this sample, we manually verified the labels of 200 randomly selected examples: 100 classified as personal narratives and 100 as non-personal narratives. This external validation showed that the classifier correctly identified personal narratives 84\% of the time (true positive rate) and non-personal narratives 97\% of the time (true negative rate), demonstrating strong overall performance. Note that this performance is considerably higher than the cross-validation metrics reported in Table 2. This is because the classifier was trained and tested on data that contained personal pronouns in every comment, making classification a harder task compared to this analysis which is on a random sample of comments which reflect our dataset's natural distribution of pronouns. 

\subsection{Identifying Politically disinclined individuals (PDIs)}
To identify PDIs, we examined two groups for each month of our analysis: users who rarely engaged in political subreddits in the previous 12 months and users who never engaged in political subreddits in the previous 12 months. 

We identified users who rarely engaged in political subreddits as follows - For each month in our analysis period (January-December 2021), we identified users who commented at least once in a political subreddit in that month but were in the bottom 25th percentile (lowest quartile) based on the number of comments they have made in political subreddits over the preceding 12 months. For example, for January 2021, we calculated the total comments made by each user who has commented in political subreddits from January 2020 to December 2020. Then, we identified the users in the lowest quartile as \textit{least politically active users}. Users in the highest quartile were considered to be the \textit{most politically active users}.

This approach, however, excludes an important group of individuals—those who were active on Reddit but not in political subreddits during the previous 12 months. We identified these users as follows - For each month, we identified users who commented at least once in a political subreddit but did not comment in political subreddits in the preceding 12 months. To avoid selecting users who were new to Reddit, we filtered out users who did not comment in any (not just political) subreddit in the preceding 12 months. We consider these users to be \textit{political newcomers}.

\section{Analysis and Results}

\subsection{H1: Using personal narratives}

H1 states that PDIs are more likely to use personal narratives than the most politically active users. To evaluate H1, for each month in our analysis period, January - December 2021, we first sampled 10,000 comments from the least politically active users, 10,000 comments from political newcomers, and 10,000 from the most politically active users. This resulted in 360,000 comments. Next, we ran the political narratives classifier on the 360,000 comments to determine which comments were personal narratives. We then conducted a random-effects logistic regression, modeling whether a comment was classified as a personal narrative or not (dependent variable). The primary independent variable of interest was the type of user (least politically active, most politically active, political newcomer). We included random effects to account for variability across subreddits and months. To account for potential confounding factors, we included the following control variables: whether the subreddit was COVID-related or not\footnote{To account for the potential impact of the pandemic on political discussions. Given the uniqueness of the pandemic, the dynamics of political discussions may have been different.}  and visibility-related measures such as whether the comment was a top-level comment in the discussion thread and cube-root transformed post score\footnote{We applied a cube-root transformation to account for the skewness in karma score since cube-root is well-defined for positive, negative and zero values unlike log transformations \cite{cox2011stata}.} (as higher scores likely imply more visibility based on Reddit's ranking algorithms). For very few comments (n=8), we did not have the post scores in the original pushshift dataset. We imputed those post scores by calculating the median (1040) of all post scores in our dataset.

The regression analyses were performed using the $lme4$ R package \cite{bates2015fitting}. The marginal means estimation and the planned contrasts to test the hypotheses were conducted using the $emmeans$ R package \cite{lenth2022emmeans}.

\begin{figure}[htbp]
    \centering
    \includegraphics[width=0.49\textwidth]{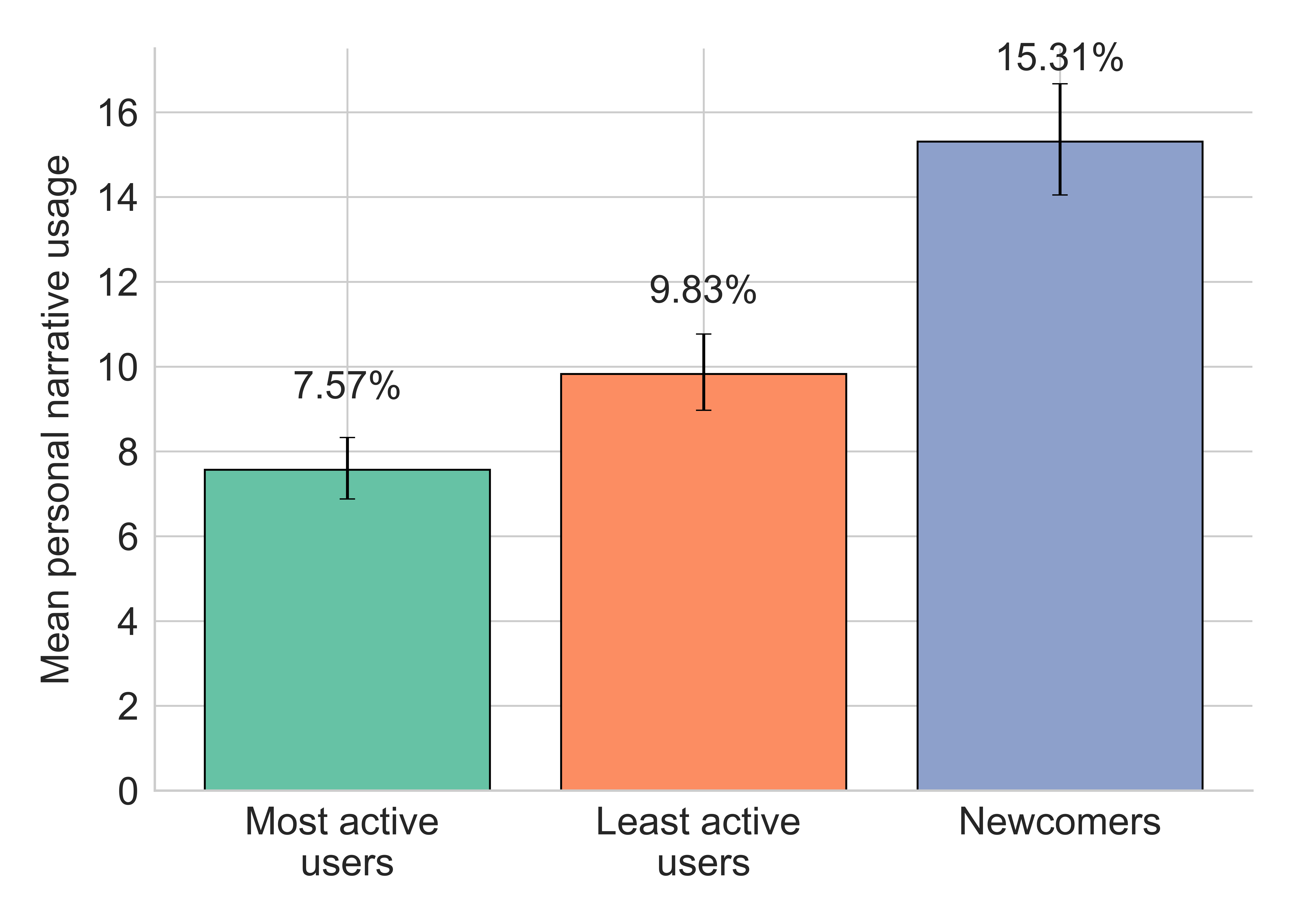} 
    \caption{Percentage use of personal narratives across user activity (95 \% CI). PDIs are more likely to use personal narratives than the most politically active users.}
    \label{Figure 1}
\end{figure}

Figure \ref{Figure 1} shows the estimated marginal mean probability of using a personal narrative for each type of user. On average, political newcomers were most likely to use personal narratives, with a probability of 15.31\% (95\% CI 14.05–16.67\%), followed by the least politically active users at 9.83\% (95\% CI 8.97–10.77\%), and finally the most politically active users at 7.57\% (95\% CI 6.88–8.33\%). The differences between these groups were statistically significant. Specifically, the least politically active users (OR = 1.33, SE = 0.025, z-ratio = 15.464, p < 0.001) and political newcomers (OR = 2.20, SE = 0.038, z-ratio = 46.206, p < 0.001) were significantly more likely to use personal narratives compared to the most politically active users. Together, these results support H1.

Interestingly, the difference between the least politically active users and political newcomers was also significant (OR = 0.60, SE = 0.009, z-ratio = -32.988, p < 0.001), suggesting that political newcomers are more inclined to rely on personal narratives to engage in political discussions than even the least politically active users. The regression table for this analysis is included in the Appendix (Table \ref{H1: Regression}).

\subsection{H2: Replying to personal narratives}

H2 states that personal narratives are more likely to receive responses from PDIs than other comments. To evaluate H2, given that we are focused on replies, we constructed a dataset by randomly sampling 20,000 comments from each month in 2021, selecting only those with at least one direct reply. This resulted in a dataset of 240,000 comments and 348,279 replies. We ran the personal narrative classifier to identify personal narratives in the 240,000 comment dataset. We identified 13,259 personal narratives (5.5\%) in this dataset. For each comment, we identified the number of political newcomers and least politically active users (PDIs) who replied to that comment. We then modeled the propensity of a PDI replying to a comment as a random-effects binomial regression. The number of unique PDIs who reply to a comment was modeled as the number of successes and the total unique users who reply to the same comment was modeled as the number of Bernoulli trials in a binomial distribution. 

The primary independent variables of interest were whether the comment being replied to was a personal narrative or not (personal narrative indicator) and user type who replied to the comment (political newcomer or least politically active user). In addition to the control variables used to evaluate H1, we included the cube-root transformed comment score as the visibility of the comment based on the comment score is likely correlated with replying behavior. We also included an interaction term between the personal narrative indicator and user type to account for potential differences in the relationships between the replying behavior of the two user groups and the use of personal narratives. Random effects were included to account for variability across subreddits and months. 

\begin{figure}[htbp]
    \centering
    \includegraphics[width=0.48\textwidth]{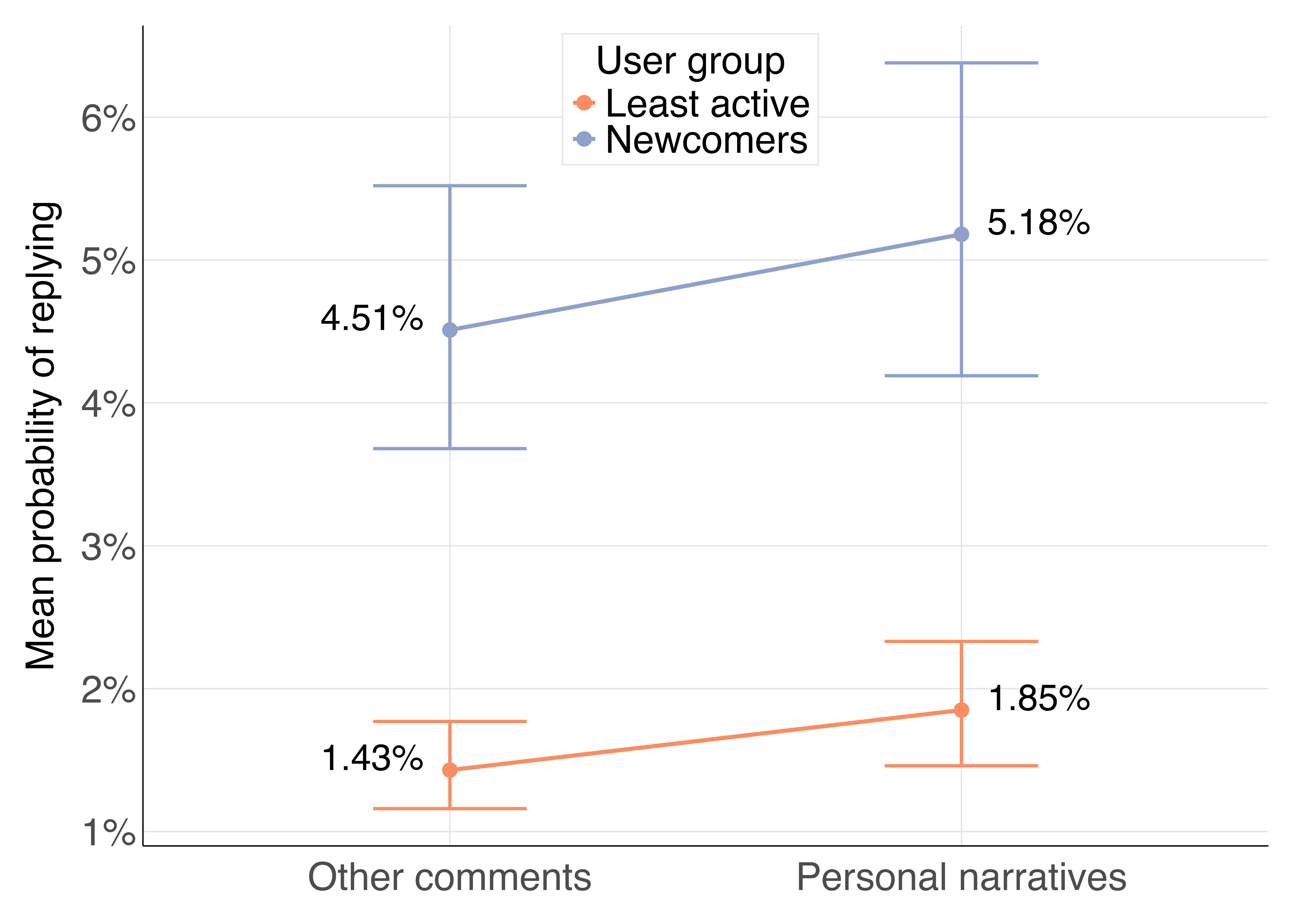} 
    \caption{Probability of receiving a reply from PDIs (95 \% CI). Personal narrative comments are slightly more likely to get a response from PDIs.}
    \label{Figure 2}
\end{figure}

Figure \ref{Figure 2} shows the estimated marginal mean probability of receiving responses from PDIs for comments with and without personal narratives. On average, comments with personal narratives (M = 1.85\%, 95\% CI 1.46–2.33\%) had a slightly higher probability of receiving a reply from the least politically active users than comments without personal narratives (M = 1.43\%, 95\% CI 1.16–1.77\%). This difference, albeit small, was statistically significant (OR = 1.30, SE = 0.075, z-ratio = 4.508, p<0.001). Similarly, comments with personal narratives (M = 5.18\%, 95\% CI 4.19–6.38\%) had a slightly higher probability of receiving a reply from political newcomers than comments without personal narratives (M = 4.51\%, 95\% CI 3.68–5.52\%). This difference, albeit small, was also statistically significant (OR = 1.16, SE=0.040, z-ratio = 4.120, p<0.001). Together, these results support H2. The regression table for this analysis is included in the Appendix (Table \ref{H2: Regression}).

\subsection{H3: Sustaining political participation}
H3 states that PDIs who use personal narratives in political communities are more likely to engage in political communities in the following month than those who do not use personal narratives. To evaluate H3, we identified if the PDIs used a personal narrative in political subreddits for each month in 2021 using the personal narrative classifier. We then checked whether these users returned to comment in political subreddits in the subsequent month. Data from December 2021 was excluded as we do not know if the users who participated in December 2021, returned in January 2022. In total, this dataset included 225,317 least politically active users and 724,096 political newcomers. 
For each unique user in each month, we identified whether they used a personal narrative in at least one of their comments during that month and whether they participated in a political discussion in the following month. We then conducted a random-effects logistic regression, modeling whether the user commented in a political subreddit in the subsequent month (dependent variable). The primary variable of interest was whether the user used a personal narrative in any of their comments in that particular month (personal narrative indicator) and the user type (least politically active and political newcomers). We also included an interaction term between the personal narrative indicator and user type to account for potential differences in the relationships between returning the subsequent month and the use of personal narratives by the two user groups. We included a random effect to account for variability across months.

\begin{figure}[htbp]
    \centering
    \includegraphics[width=0.48\textwidth]{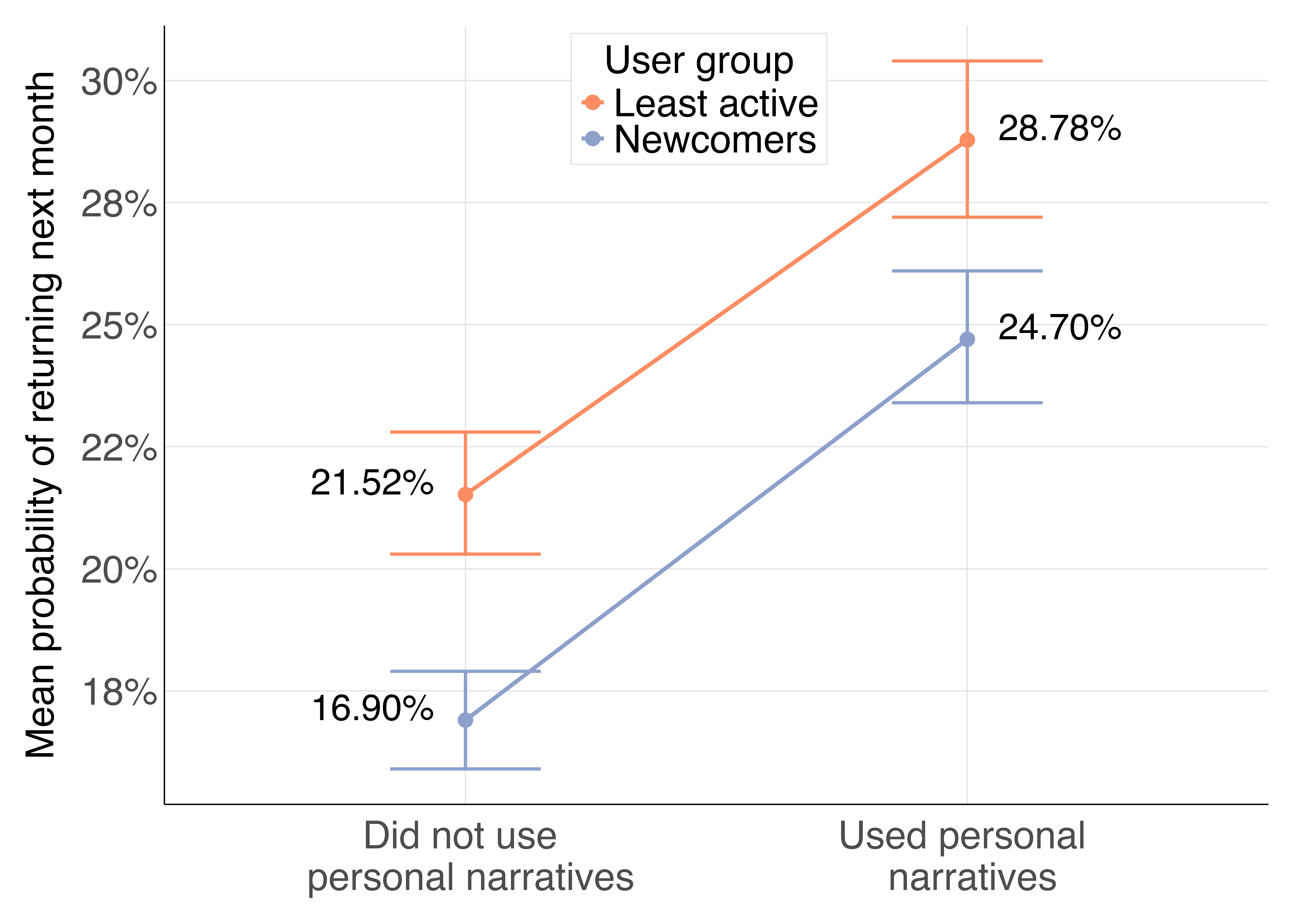} 
    \caption{Probability of returning next month to political discussions for PDIs (95\% CI). PDIs are more likely to return to political discussions in the next month if they employ personal narratives in the current month.}
    \label{Figure 3}
\end{figure}

Figure~\ref{Figure 3} shows the average probability of the least politically active users and political newcomers returning to political subreddits the following month, based on whether they used a personal narrative in their comments in a particular month. On average, the least politically active users who used personal narratives in a given month ($M = 28.78\%$, 95\% CI $27.24\%$–$30.37\%$) were more likely to engage in political discussions in the subsequent month compared to those who did not use personal narratives ($M = 21.52\%$, 95\% CI $20.32\%$–$22.76\%$). This difference was statistically significant (OR = $1.47$, SE = $0.022$, $z$-ratio = $26.26$, $p < 0.001$). Similarly, on average, political newcomers who used personal narratives in a given month ($M = 24.70\%$, 95\% CI $23.36\%$–$26.07\%$) were more likely to engage in political discussions in the subsequent month than those who did not use personal narratives ($M = 16.90\%$, 95\% CI $15.91\%$–$17.93\%$). This difference was statistically significant (OR = $1.613$, SE = $0.012$, $z$-ratio = $64.294$, $p < 0.001$). Together, these results support H3. The regression table for this analysis is included in the Appendix (Table \ref{H3: Regression}).

\subsection{H4, H5, RQ1: Community perception of personal narratives}
H4, H5 and RQ1 evaluate the perceptions of using personal narratives. H4 states that comments by PDIs are likely to score lower than comments by the most politically active users. H5 states that personal narrative comments by PDIs are more likely to score higher than other comments by PDIs. RQ1 asks how the score of personal narrative comments made by PDIs and the most politically active users vary. 

To evaluate these hypotheses and the research question, we use the same dataset used to evaluate H1. We used a linear regression model to model the cube-root transformed score of the comment as the dependent variable. We include the type of user posting the comment (political newcomers, least politically active, most politically active) and whether the comment was a personal narrative or not (personal narrative indicator) as independent variables. We also included an interaction term between user type and the personal narrative indicator to model how the relationship between different user types and the cube-root transformed comment score varies based on whether a comment is a personal narrative or not. To account for potential confounding factors, we included the same control variables we used to evaluate H1. Random effects were included to account for variability across subreddits and months. 

\begin{figure}[htbp]
    \centering
    \includegraphics[width=0.48\textwidth]{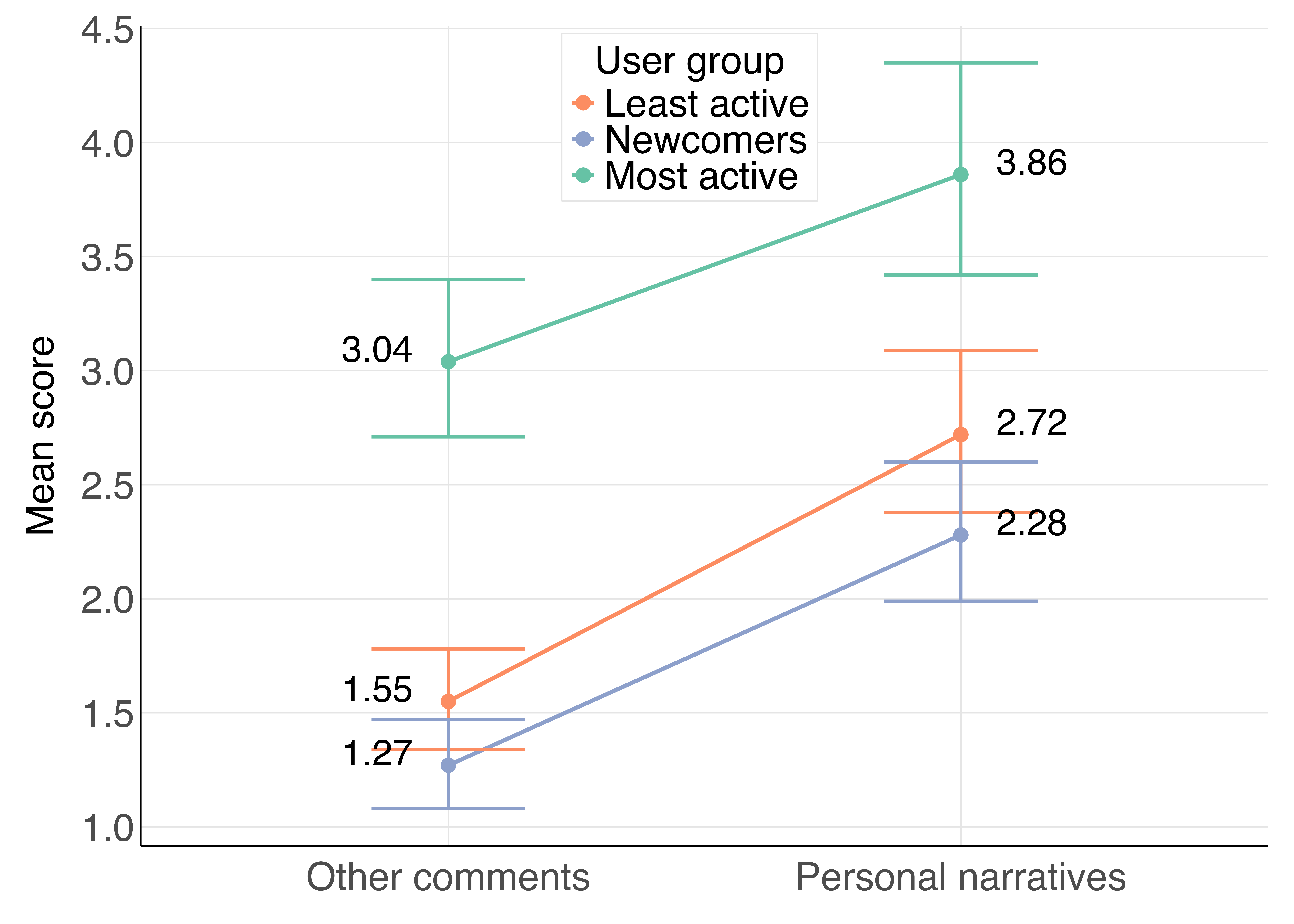} 
    \caption{Mean comment scores of users across different comment types (95\% CI). PDIs are likely to score lower than comments by the most politically active users. Also, personal narrative comments by PDIs are likely to score higher than other comments made by the same users.}
    \label{Figure 4}
\end{figure}

Figure \ref{Figure 4} shows the estimated mean comment scores of comments across users with different political activity levels (least politically active vs.\ most politically active vs.\ political newcomers) and comment types (comments with personal narrative vs.\ other comments). We observe that comments made by the least politically active users and political newcomers had lower estimated mean scores than those made by the most politically active users for both comments containing a personal narrative (least politically active users: $M = 2.72$, 95\% CI $2.38$–$3.09$, political newcomers: $M = 2.28$, 95\% CI $1.99$–$2.60$, most politically active users: $M = 3.86$, 95\% CI $3.42$–$4.35$) and those that do not (least politically active users: $M = 1.55$, 95\% CI $1.34$–$1.78$, political newcomers: $M = 1.27$, 95\% CI $1.08$–$1.47$, most politically active users: $M = 3.04$, 95\% CI $2.71$–$3.40$). These differences were statistically significant (see Appendix table \ref{tab:contrasts} for details). Together, these results support H4. The differences between mean scores of comments containing personal narratives and those that do not, authored by the least politically active users (M = $0.24$, SE = $0.0148$, $z$-ratio = $16.26$, $p < 0.001$) and political newcomers (M = $0.23$, SE = $0.0119$, $z$-ratio = $19.62$, $p < 0.001$) are also statistically significant. These results support H5. The regression table for this analysis is included in the Appendix (Table \ref{RQ: Regression}).

Further, answering RQ1, we compare the difference between scores of personal narrative comments by PDIs and the most politically active users with the difference between the scores of other comments by the same group of users. We find that, on average, the difference in mean scores between the least politically active users and the most politically active users is significantly smaller when both groups use personal narratives compared to when they do not (difference in cube-root scale: $0.119$, SE = $0.022$, $z$-ratio = $5.320$, $p <0.001$). We observe similar smaller differences between political newcomers and the most politically active users as well when both groups use personal narratives (difference in cube-root scale: $0.113$, SE = $0.021$, $z$-ratio = $5.512$, $p <0.001$). These results indicate that personal narratives could help reduce the disparity in positive reception between content posted by PDIs and the most politically active users. 

\section{Discussion}

The results of this study indicate that personal narratives are a promising way to encourage politically disinclined individuals to engage in political discussions.  We find that PDIs are more likely to use personal narratives (H1) and are slightly more likely to respond to them (H2). Further, PDIs who use personal narratives are more likely to return to future political discussions (H3). Also, personal narratives appear to positively contribute to political discussions as well. We find that comments by PDIs containing personal narratives are more likely to be viewed positively than other comments by them (H5). Although comments by PDIs, on average, score lower points than those by more politically active users (H4), personal narratives appear to reduce this difference in scores (RQ1). Overall, these results highlight that personal narratives can positively impact both participation in and the overall quality of online political discussions. Below, we discuss the implications of our findings.

Support for H1 and H2 highlights the inclusivity and accessibility of personal narratives as a form of communication. As discussed earlier, many individuals avoid engaging in political discussions because they do not know enough about the issue \cite{carlson2022goes}. These results suggest that personal narratives provide a low-barrier entry point into political discussions, reducing the reliance on prior political knowledge or expertise. Yet traditionally, political discussion spaces privilege rational arguments over personal stories and testimonials \cite{young2002inclusion}. Similarly, many online political communities restrict the use of personal narratives. For example, the rules in the r/neutralpolitics community explicitly state that ``if you're claiming something to be true, you need to back it up with a qualified source. There is no `common knowledge' exception, and anecdotal evidence is not allowed.'' Our results suggest that such an approach to content moderation, while facilitating a fact-based discussion space, likely has an unintended effect of turning away individuals who do not typically engage in political discussions.

To survive, all communities, especially political ones, need to attract and retain newcomers in order to counter inevitable user churn and burnout \cite{kraut2012building}.  Our results suggest that community moderators can encourage the use of personal narratives to attract and retain new users into the community. This can be done through norm-setting by highlighting exemplar comments containing personal narratives and by explicitly noting their usage in the community rules.

Attracting and retaining users is only one part of the puzzle for content moderators. While newcomers often provide fresh ideas and help stimulate discussions in newer directions \cite{levine2003newcomer}, their contributions may not always align with the norms of the communities \cite{kiesler2012regulating}. Indeed, in our analysis, political communities perceive the contributions of the PDIs less positively than those of the more politically active users. However, we find that the gap in community perceptions (based on karma score) of politically disinclined and more politically active users reduces when PDIs employ personal narratives. This may be achieved by encouraging introductory posts by newcomers to share their personal stories. Further, given that personal narratives also generally score higher than other comments, this early positive feedback may provide additional encouragement to newcomers to productively engage in the community. As Kraut et al. \cite{kraut2012building} note, ``when newcomers have friendly interactions with existing community members soon after joining a community, they are more likely to stay longer and contribute more.''

While these results are promising, their potential to drive actionable change can be fully unlocked only when combined with implementable strategies that enhance political discourse in online communities. Social media platforms, in particular, hold the power to amplify the visibility of personal narratives. For example, Bail highlights how algorithms can be optimized to prioritize content with broad resonance, rather than divisive or controversy-driven engagement metrics \cite{bail2022breaking}. Similarly, platforms could adapt their algorithms to elevate content featuring personal narratives, helping to foster a more inviting environment that encourages participation by a wider range of users. Alternately, platforms can leverage gamification such as badges for first-time contributors who employ personal narratives, to encourage disinclined users to share personal narratives in political discussion spaces. 

Finally, political scientists have flagged concerns about how online political discussions are being dominated by users who are heavily politically involved, particularly how this often leads to a hostile partisan environment and misperceptions about polarization \cite{krupnikov2022other}. In an ideal scenario, a broader cross-section of individuals with differing political involvement participate in these discussions. Our results suggest that encouraging the use of personal narratives in political discussions likely attracts and retains users who typically avoid politics, advancing the normatively positive goal of broadening political participation online.

\section{Limitations and Future Work}

While our study demonstrates the potential of personal narratives to engage politically disinclined individuals, we acknowledge its limitations. The research relied on Reddit comments from political subreddits, which may not fully represent broader online political discourse due to platform-specific demographic biases. Moreover, the dataset's temporal constraints---limited to 2020-2021---coincided with the COVID-19 pandemic and heightened political polarization, potentially amplifying narrative trends that might differ under more typical conditions. The study's definition of political activity, based solely on subreddit engagement, might not be complex enough and could alter the findings. Finally, external events in the users life might prompt users to both participate in a political subreddit and share their personal experiences. Thus, sharing personal experiences may be issue specific and may not extend to other topics. Future research can address atleast some of these limitations by expanding the investigative scope across multiple platforms, conducting longitudinal studies, and incorporating broader definitions of political engagement. 

While this study primarily highlights the positive effects of personal narratives, they can also be used to mislead and spread misinformation, particularly in political discussions \cite{van2021speaking}. We do not investigate whether personal narratives were employed for misinformation in this study, leaving it as a potential avenue for future research. Additionally, we do not analyze the content of personal narratives in this work.

Future potential research directions include comparing narrative patterns across social media platforms like Twitter and Facebook, examining the persistence of engagement facilitated by personal narratives, and exploring their potential impact on tangible political outcomes such as voter turnout and activism. We hope that this study can catalyze future research on how personal narratives might bridge the gap between political apathy and active participation, ultimately ensuring more inclusive and accessible political discourse.

\begin{acks}
The authors acknowledge the generous support from the Center for Media Engagement and the Moody College of Communication at The University of Texas at Austin. The authors also thank Utkarsh Mujumdar and Chris Hickey for their early assistance on this project. 
\end{acks}

\bibliographystyle{ACM-Reference-Format}
\bibliography{main}

\appendix

\section{Appendix: Regression Tables}

\begin{table*}[t]
\begin{minipage}{0.48\textwidth}
\centering 
\caption{Regression coefficients for H1} 
\label{H1: Regression} 
\small 
\begin{tabular}{@{\extracolsep{0pt}}lc} 
\\[-1.8ex]\hline 
\hline \\[-1.8ex] 
 & \multicolumn{1}{c}{\textit{Dependent variable:}} \\ 
\cline{2-2} 
\\[-1.8ex] & Personal narrative or not \\ 
\hline \\[-1.8ex] 
\textbf{User type (vs most active users)} & \\ 
\\[-1.8ex] 
Least politically active & 0.286$^{***}$ (0.018) \\ 
Political newcomers & 0.791$^{***}$ (0.017) \\ 
\\ 
\textbf{Control variables} & \\ 
\\[-1.8ex] 
Is top-level comment & $-$0.215$^{***}$ (0.015) \\ 
COVID-19 subreddit & 1.426$^{***}$ (0.090) \\ 
Post score (cuberoot) & 0.008$^{***}$ (0.001) \\ 
\\ 
Constant & $-$3.214$^{***}$ (0.045) \\ 
\hline \\[-1.8ex] 
Observations & 360,000 \\ 
Log Likelihood & $-$88,302.770 \\ 
\hline 
\hline \\[-1.8ex] 
\textit{Note:}  & \multicolumn{1}{r}{$^{*}$p$<$0.1; $^{**}$p$<$0.05; $^{***}$p$<$0.01} \\ 
\end{tabular}
\end{minipage}
\hfill
\begin{minipage}{0.48\textwidth}
\centering 
\caption{Regression coefficients for H2} 
\label{H2: Regression} 
\small 
\begin{tabular}{@{\extracolsep{0pt}}lc} 
\\[-1.8ex]\hline 
\hline \\[-1.8ex] 
 & \multicolumn{1}{c}{\textit{Dependent variable:}} \\ 
\cline{2-2} 
\\[-1.8ex] & Participation next month \\ 
\hline \\[-1.8ex] 
\textbf{Main effects} & \\ 
\\[-1.8ex] 
Used personal narrative & 0.261$^{***}$ (0.058) \\ 
Political newcomers (vs least active) & 1.179$^{***}$ (0.019) \\ 
\\ 
\textbf{Control variables} & \\ 
\\[-1.8ex] 
Is top-level comment & 0.022 (0.018) \\ 
COVID-19 subreddit & 0.378$^{***}$ (0.139) \\ 
Post score (cuberoot) & 0.017$^{***}$ (0.001) \\ 
Comment score (cuberoot) & 0.013$^{***}$ (0.003) \\ 
\\ 
\textbf{Interaction term} & \\ 
\\[-1.8ex] 
Used personal narrative $\times$ newcomers & $-$0.115$^{*}$ (0.067) \\ 
\\ 
Constant & $-$4.638$^{***}$ (0.099) \\ 
\hline \\[-1.8ex] 
Observations & 480,000 \\ 
Log Likelihood & $-$66,425.730 \\ 
\hline 
\hline \\[-1.8ex] 
\textit{Note:}  & \multicolumn{1}{r}{$^{*}$p$<$0.1; $^{**}$p$<$0.05; $^{***}$p$<$0.01} \\ 
\end{tabular}
\end{minipage}
\end{table*}

\begin{table*}[t]
\begin{minipage}{0.48\textwidth}
\centering 
\caption{Regression coefficients for H3} 
\label{H3: Regression} 
\small 
\begin{tabular}{@{\extracolsep{0pt}}lc} 
\\[-1.8ex]\hline 
\hline \\[-1.8ex] 
 & \multicolumn{1}{c}{\textit{Dependent variable:}} \\ 
\cline{2-2} 
\\[-1.8ex] & User active next month \\ 
\hline \\[-1.8ex] 
\textbf{Main effects} & \\ 
\\[-1.8ex] 
Used personal narrative & 0.388$^{***}$ (0.015) \\ 
Political newcomers (vs least active)  & $-$0.299$^{***}$ (0.006) \\ 
\\ 
\textbf{Interaction term} & \\ 
\\[-1.8ex] 
Used personal narrative $\times$ newcomers & 0.090$^{***}$ (0.017) \\ 
\\ 
Constant & $-$1.294$^{***}$ (0.037) \\ 
\hline \\[-1.8ex] 
Observations & 949,413 \\ 
Log Likelihood & $-$459,251.000 \\ 
\hline 
\hline \\[-1.8ex] 
\textit{Note:}  & \multicolumn{1}{r}{$^{*}$p$<$0.1; $^{**}$p$<$0.05; $^{***}$p$<$0.01} \\ 
\end{tabular}
\end{minipage}
\hfill
\begin{minipage}{0.48\textwidth}
\centering 
\caption{Regression coefficients for H4 and H5} 
\label{RQ: Regression} 
\small 
\begin{tabular}{@{\extracolsep{5pt}}lc} 
\\[-1.8ex]\hline 
\hline \\[-1.8ex] 
 & \multicolumn{1}{c}{\textit{Dependent variable:}} \\ 
\cline{2-2} 
\\[-1.8ex] & Transformed comment score \\ 
\hline \\[-1.8ex] 
\textbf{User type (vs most active users)} & \\ 
\\[-1.8ex] 
Least politically active  & $-$0.292$^{***}$ (0.005) \\ 
Political newcomers & $-$0.367$^{***}$ (0.006) \\ 
\\ 
\textbf{Interaction terms} & \\ 
\\[-1.8ex] 
Used personal narrative $\times$ least active & 0.118$^{***}$ (0.022) \\ 
Used personal narrative $\times$ newcomers & 0.135$^{***}$ (0.021) \\ 
\\ 
\textbf{Control variables} & \\ 
\\[-1.8ex] 
Is top-level comment & 0.078$^{***}$ (0.005) \\ 
COVID-19 subreddit & 0.195$^{**}$ (0.042) \\ 
Post score (cuberoot) & $-$0.002$^{***}$ (0.000) \\ 
\\ 
Constant & 1.379$^{***}$ (0.024) \\ 
\hline \\[-1.8ex] 
Observations & 360,000 \\ 
\hline 
\hline \\[-1.8ex] 
\textit{Note:}  & \multicolumn{1}{r}{$^{*}$p$<$0.1; $^{**}$p$<$0.05; $^{***}$p$<$0.01} \\ 
\end{tabular}
\end{minipage}
\end{table*}

\begin{table*}[htbp]
\centering
\caption{Contrasts for H4}
\label{tab:contrasts}
\begin{tabular}{|c|c|c|c|c|c|}
\hline
\textbf{Comment type} & \textbf{Comparisons}& \textbf{Estimate} & \textbf{SE} & \textbf{z-ratio} & \textbf{p-value} \\
\hline
{Personal narrative} 
& Most politically active - least politically active  & 0.1733 & 0.02171 & 7.980 & $<.0001$ \\ \cline{2-6}
& Most politically active - political newcomers & 0.2530 & 0.01992 & 12.699 & $<.0001$ \\ \hline
{Other comments} 
& Most politically active - least politically active & 0.2920 & 0.00540 & 54.074 & $<.0001$ \\ \cline{2-6}
& Most politically active - political newcomers  & 0.3665 & 0.00545 & 67.192 & $<.0001$ \\ 
\hline
\end{tabular}
\end{table*}

\end{document}